\documentclass[twocolumn,prb,citeautoscript,superscriptaddress,8pt]{revtex4-1}

\usepackage{graphicx}
\usepackage{gensymb}
\usepackage{color}
\usepackage[dvipsnames]{xcolor}
\usepackage{float}
\usepackage{upgreek}
\usepackage{bm}
\usepackage{amsmath,amssymb}
\newcommand{\ket}[1]{\ensuremath{\left| #1 \right\rangle}}

\begin{document}

\title{Imaging current control of magnetization in Fe$_3$GeTe$_2$ with a widefield nitrogen-vacancy microscope}

\author{Islay O. Robertson}
\affiliation{School of Science, RMIT University, Melbourne, VIC 3001, Australia}
\affiliation{School of Physics, University of Melbourne, VIC 3010, Australia}

\author{Cheng Tan}
\affiliation{School of Science, RMIT University, Melbourne, VIC 3001, Australia}

\author{Sam C. Scholten}
\affiliation{School of Physics, University of Melbourne, VIC 3010, Australia}
\affiliation{Centre for Quantum Computation and Communication Technology, School of Physics, University of Melbourne, VIC 3010, Australia}

\author{Alexander J. Healey}
\affiliation{School of Physics, University of Melbourne, VIC 3010, Australia}
\affiliation{Centre for Quantum Computation and Communication Technology, School of Physics, University of Melbourne, VIC 3010, Australia}

\author{Gabriel J. Abrahams}
\affiliation{School of Science, RMIT University, Melbourne, VIC 3001, Australia}
\affiliation{School of Physics, University of Melbourne, VIC 3010, Australia}

\author{Guolin Zheng}
\affiliation{School of Science, RMIT University, Melbourne, VIC 3001, Australia}

\author{Aur{\'e}lien Manchon}
\affiliation{CINaM, Aix-Marseille Université, CNRS, Marseille, France}

\author{Lan Wang}
\affiliation{School of Science, RMIT University, Melbourne, VIC 3001, Australia}

\author{Jean-Philippe Tetienne}
\email{jean-philippe.tetienne@rmit.edu.au}
\affiliation{School of Science, RMIT University, Melbourne, VIC 3001, Australia}

\begin{abstract} 

Van der Waals (vdW) magnets are appealing candidates for realising spintronic devices that exploit current control of magnetization (e.g. switching or domain wall motion), but so far experimental demonstrations have been sparse,
in part because of challenges associated with imaging the magnetization in these systems. Widefield nitrogen-vacancy (NV) microscopy allows rapid, quantitative magnetic imaging across entire vdW flakes, ideal for capturing changes in the micromagnetic structure due to an electric current. Here we use a widefield NV microscope to study the effect of current injection in thin flakes ($\sim10$\,nm) of the vdW ferromagnet Fe$_3$GeTe$_2$ (FGT). We first observe current-reduced coercivity on an individual domain level, where current injection in FGT causes substantial reduction in the magnetic field required to locally reverse the magnetisation. We then explore the possibility of current-induced domain-wall motion, and provide preliminary evidence for such a motion under relatively low current densities, suggesting the existence of strong current-induced torques in our devices. Our results illustrate the applicability of widefield NV microscopy to imaging spintronic phenomena in vdW magnets, highlight the possibility of efficient magnetization control by direct current injection without assistance from an adjacent conductor, and motivate further investigations of the effect of currents in FGT and other vdW magnets. 

\end{abstract}

\maketitle 

\section{Introduction}

Exfoliation of two-dimensional (2D) van der Waals (vdW) magnets has created opportunities for developing new methods for integrating magnetic materials into future and existing technologies \cite{Cortie2020}. Their low dimensionality favours the use of weak external stimuli such as strain, electric fields, or low-power currents to alter or interact with the micromagnetic structure \cite{Wenbin2021,Kurebayashi2022}. Furthermore, their freestanding nature facilitates stacking layers of material at different angles to create twisted structures or layering with different materials (heterostructures) to create new interface interactions \cite{Gibertini2019,Gong2019}. Exploiting external stimuli and interface interactions are driving the development of new devices based on vdW magnets \cite{Klein2018,Zhe2018,Albarakati2019,Kaixuan2021, KaixuanAFM2021,Gong2019,Sierra2021}.
An area of significant interest for 2D vdW magnets is spintronics, where external stimulation and interface interactions play an important role in creating a variety of phenomena \cite{Sierra2021,Xiaoyang2019,Anjan2016,Manchon2019}, which primarily involve the interaction between the magnetic structure of a host material and the magnetic moment of injected electrons. Current injection will generally either alter the behaviour of the injected electrons, or induce changes in the micromagnetic structure of the host material. A particular phenomenon which changes the domain structure of the host magnetic material is current-induced domain-wall motion (CIDM) where the injection of currents into the material (or in an adjacent conductor) causes growth/shrinkage of adjacent magnetic domains thereby moving the domain wall separating them \cite{MironNM2011,Emori2013,Ryu2013}, which can be extended to more complex magnetic textures such as skyrmions \cite{Fert2017}. Significant interest in CIDM is primarily due to the mechanisms driving the motion, namely the spin-transfer torque (STT) and spin-orbit torque (SOT) which are often studied in conventional magnetic multilayer systems for potential applications in future highly efficient magnetic memory and logic devices \cite{Manchon2019}. Note that while such multilayer systems are generally ferromagnetic with perpendicular magnetic anisotropy, antiferromagnets, synthetic or intrinsic, can be more efficient for CIDM than their ferromagnetic counterparts due to the absence of stray fields stabilizing adjacent domains, however, this also makes them exceptionally difficult to study \cite{Manchon2019}.
%and exhibit perpendicular magnetic anisotropy, though this can limit the efficiency of CIDM as the stray fields stabilizes adjacent domains. Antiferromagnets, synthetic or intrinsic, remove the stray fields through spin compensation and thus can be more efficient than their ferromagnetic counterparts however, this also makes them exceptionally difficult to study \cite{Manchon2019}}.
Identification of SOTs in vdW systems offers new pathways for engineering devices with the potential for even greater efficiencies \cite{Wei2021,Hyunsoo2022}. While magnetization switching by SOT has been achieved in various 2D or hybrid 2D/3D systems \cite{Alghamdi2019,Wang2019,Gupta2020,Inseob2022}, so far CIDM has not been observed in vdW structures \cite{Abdul2021}. 

Commonly, spintronic phenomena are inferred from transport measurements, which give little insight into the micromagnetic structure as they offer no means of direct observation of the material \cite{Mak2019}. To fully understand changes in the domains, e.g. to unambiguously identify CIDM, high resolution magnetic imaging techniques need to be employed. 
Magnetic imaging is most often applied to the study of 2D vdW magnets using scanning probe techniques and magneto-optical effects. Scanning probe techniques, e.g. based on single nitrogen-vacancy (NV) centers in diamond, often yield the highest spatial resolution at the cost of longer image collection times and smaller fields of view, limiting images to small sections of the subject material \cite{Thiel2019}. Magneto-optical methods based on the Kerr effect and magnetic circular dichroism are capable of larger fields of view but materials must facilitate the necessary optical interactions and require careful calibration to make quantitative measurements \cite{Mak2019}. Such restrictions can inhibit the study of spintronic phenomena by preventing quantitative analysis and limiting studies to certain materials or designated regions.  
To compensate for some of the shortcomings of preexisting magnetic imaging techniques, here we utilise widefield NV microscopy \cite{Levine2019,Scholten2021} as a method for imaging 2D vdW magnets, by optically probing magnetic stray fields from a sample material via a proximal, dense layer of NV centers. Widefield NV microscopy, also known as quantum diamond microscopy, offers calibration-free quantitative imaging with sub-micrometer resolution over a comparatively large field of view, and high measurement throughput, properties which suitably lend themselves to systematically studying the micromagnetic structure across the full surface of 2D vdW magnets, as demonstrated recently by several groups \cite{BroadwayAM2020,McLaughlin2021,Hang2022}. Similarly, these features are also desirable for imaging spintronic phenomena \cite{Yan2022}, but the technique is yet to be applied to vdW spintronic devices. Here, we apply widefield NV microscopy to studying current-induced changes in the micromagnetic structure of a 2D vdW ferromagnet using Fe$_3$GeTe$_2$ (FGT) \cite{Deiseroth2006,Tan2018,Zaiyao2018,Yujun2018} as the host material. 

FGT stands out amongst other vdW magnets as it is a ferromagnetic metal with large anomalous Hall conductance \cite{Kyoo2018} which, along with its conductivity and relatively high Curie temperature ($\approx200$\,K \cite{Tan2018}), makes it ideal for studying spintronic phenomena. Additionally, its crystal symmetry supports strong spin-orbit coupling which can stabilise chiral spin structures and skyrmions \cite{Ding2020, Hong2020, ChakrabortyAM2022}. A number of current-induced spintronic effects have been identified in FGT including: current-reduced coercivity \cite{Kaixuan2021}, antisymmetric magnetoresistance \cite{Albarakati2019}, bilayer-assisted magnetization switching \cite{Wang2019, Alghamdi2019,Inseob2022}, and skyrmion motion \cite{TaeEon2021}. Here, we aim to study the possibility of CIDM in FGT thin flakes when the current is injected directly into the FGT layer, without assistance from an adjacent conductive layer. Our study was inspired by recent works \cite{Johansen2019,Martin2021,Kaixuan2021} which suggested an efficient bulk SOT in FGT by direct current injection, an intriguing departure from the bilayer systems mainly studied so far, in which the current flows in a non-magnetic conductor creating an interfacial SOT acting on the adjacent ferromagnet  \cite{Emori2013,Ryu2013,Manchon2019,Alghamdi2019,Wang2019,Gupta2020,Inseob2022}. We first use our microscope to characterise FGT flakes without electrodes, and highlight the requirements for imaging contacted FGT devices. We then image such a device, observe a local reduction in coercivity due to current pulses, and suggest some ideal conditions for observing CIDM. Finally we provide the evidence of CIDM in a separate FGT device, at relatively low current densities. The possible nature of the magnetisation dynamics and role of current-induced torques (e.g. SOT) are discussed. Our results suggest the possibility of efficient control of the magnetic microstructure of thin FGT flakes by direct current injection, and motivate further investigations of the effect of currents in FGT and other vdW magnets. They also illustrate the applicability, and associated challenges, of widefield NV microscopy to imaging spintronic phenomena in vdW magnets.

\section{Results and Discussion}

\begin{figure*}[tb]
\centering
\includegraphics[width=0.9\textwidth]{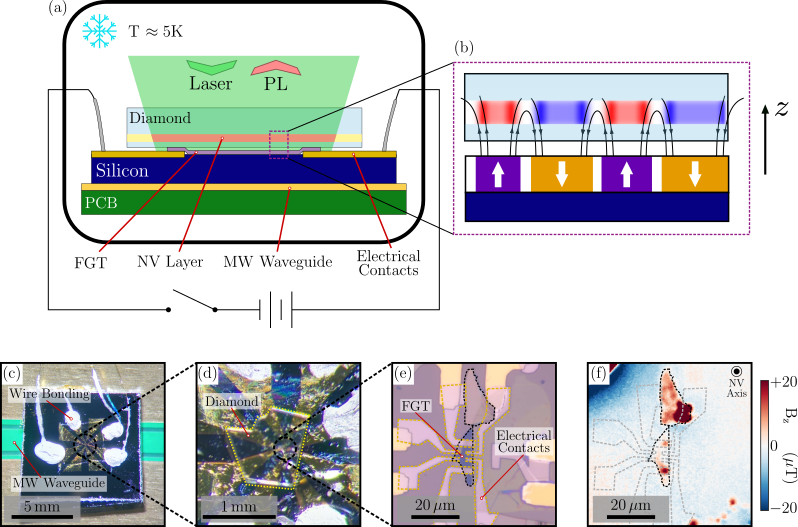}
\caption{{\bf Imaging spintronic devices with widefield NV microscopy.} (a) Cross-sectional schematic of a spintronic device with a surface mounted diamond containing a shallow nitrogen-vacancy (NV) layer. The device centrepiece is a thin flake of Fe$_3$GeTe$_2$ (FGT), a van der Waals magnet which conforms to the silicon substrate/electrical contact topography. The device is placed on a printed circuit board (PCB) with a microwave (MW) waveguide for driving the NV spin resonances. For measurement the entire device is placed in a cryogenic chamber with a base temperature of $\approx 5$\,K. (b) Stray fields from the domain structure are mapped by optically interrogating the array of NVs which act as local magnetometers. In this way a map of the stray field at the NV layer is measured. (c) Optical image of an example spintronic device mounted on a circuit board and connected to a power source via wire bonding for current injection. (d) Optical micrograph of the device in (c) with a diamond mounted on top. The diamond is glued at its corners to ensure it remains in place when loading into the cryostat to help maintain close proximity to the device. (e) Optical micrograph at the centre of the spintronic device (taken prior to diamond mounting) showing flakes of FGT (black contours) positioned over platinum contacts (yellow contours). (f) Stray field image of the device in (e) obtained with widefield NV microscopy, recorded under a bias field $B_{\rm NV}^0 = 6$\,mT applied along the NV axis coinciding with the $z$ axis. Prior to imaging, the device was heated to near the Curie temperature and then cooled under the $6$\,mT field.}
\label{fig1}
\end{figure*}

Widefield NV microscopy employs a diamond sensor which is interfaced with the sample of interest \cite{Scholten2021}, in this instance a contacted spintronic device [Fig.~\ref{fig1}(a)]. The sensor is composed of a shallow layer of magnetically sensitive NVs embedded in the diamond lattice whose ground state spin transitions are influenced by stray fields emanating from a proximal magnetic material such as FGT [Fig.~\ref{fig1}(b)]. Illuminating the NV layer with a green ($532$\,nm) laser, coupled with microwave excitation, probes the NV spin transitions which are then readout by their red photoluminesence in a process known as optically detected magnetic resonance (ODMR) \cite{Doherty2013,Rondin2014,Casola2018}. 
To study spintronic phenomena, we constructed devices by exfoliating flakes of FGT (thickness $\sim10$\,nm) and transferring them onto platinum contacts which were fabricated on a silicon substrate and then wire-bonded to an external power source. The device is mounted on a printed circuit board which contains a coplanar microwave waveguide for driving the NVs' spin transitions [Fig.~\ref{fig1}(c)]. A diamond is then placed onto the surface of the device in such a way to ensure the NV layer is in close contact with the flake of FGT [Fig.~\ref{fig1}(d)] (method in SI, Sec. III). The FGT flake bridges a number of contacts [Fig.~\ref{fig1}(e)] and is coated with a protective layer to limit atmospheric exposure (details in SI, Sec. II). 

\begin{figure*}[tb]
\centering
\includegraphics[width=0.9\textwidth]{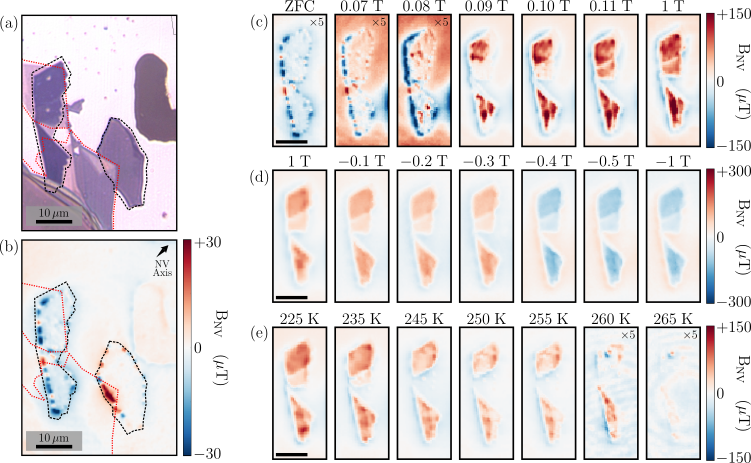}
\caption{{\bf Imaging the ferromagnetic properties of FGT flakes.} (a) Flakes of FGT (black outline) (maximum thickness $18$\,nm or 23 layers) on the surface of an Al-coated diamond. FGT is placed on top of flakes of WTe$_2$ (red outline) ($\approx 3$\,nm or 3 layers) with partial overlap with FGT forming heterostructures locally. The features not outlined are registration marks made in the Al coating. (b) Stray-field image of the FGT flakes after zero field cooling, recorded under a bias field $B_{\rm NV}^0 = 6$\,mT applied along the NV axis (direction indicated by the arrow in the top-right corner, pointing $55^\circ$ off the $z$ axis). (c-e) Stray-field image series for probing the depinning field c), switching field d), and temperature dependent phase transition e) in FGT for one of the flakes in a). (c) To measure the depinning field, the series starts from the zero-field-cooled state and then images are taken after fields of increasing magnitude are applied along the easy axis of magnetization ($z$ axis) until the flake is observed to be fully magnetized ($\approx 0.11$\,T). (d) For the switching field, fields of increasing magnitude are applied in opposition  to a positive single domain state until there is a reversal of polarity in the stray field. (e) The temperature dependent phase transition can be studied by imaging under increasing temperatures. Starting at $225$\,K the temperature is increased until stray fields are no longer observed from the flake at $265$\,K. All images are taken under a bias field $B_{\rm NV}^0 = 6$\,mT.}
\label{fig2}
\end{figure*}

An example magnetic image for the device in Fig.~\ref{fig1}(e) taken with the widefield NV microscope shows magnetic fields emanating almost exclusively from the outlined flakes of FGT [Fig.~\ref{fig1}(f)]. The image is generated by simultaneously recording ODMR spectra at each pixel on a camera. To extract stray field information from the ODMR spectra, images are taken under a bias field ($B_{\rm NV}^0 = 6$\,mT) applied along a particular NV axis creating two ODMR peaks separated by $\Delta f = 2\gamma_e(B_{\rm NV}^0 + B_{\rm NV})$ where $\gamma_e\approx28$\,GHz\,T$^{-1}$ is the electron gyromagnetic ratio \cite{Rondin2014}. In this way the projection of the stray field from the material along the specified NV axis ($B_{\rm NV}$) can be recovered \cite{Scholten2021}. We collect our magnetic images using a custom-built widefield microscope system constructed around a cryogenic chamber \cite{Lillie2020} to facilitate imaging below the Curie temperature of FGT. Except in Fig.~\ref{fig2}, the diamond crystal orientation is chosen so the projection axis coincides with the out-of-plane ($z$) axis, i.e.\,$B_{\rm NV}=B_z$. Given FGT is magnetized out of plane due to perpendicular magnetic anisotropy \cite{Tan2018}, there is a relatively simple correspondence between measured stray field and underlying magnetization, as illustrated in Fig.~\ref{fig1}(b), which facilitates visualisation of magnetic domains.

\subsection*{Characterisation of isolated FGT flakes}

As a first step towards CIDM, we characterised the ferromagnetic properties of isolated FGT flakes (i.e.\,without electrical contacts) using the widefield NV microscope. In this way we can build an understanding of the conditions under which we will be trying to move domain walls. For this characterisation, flakes of FGT were exfoliated from bulk material and transferred directly onto the surface of a diamond [Fig.~\ref{fig2}(a)]. Doing so allows us to probe field and temperature dependence of the magnetic properties with optically limited spatial resolution \cite{BroadwayAM2020}. Under a bias field of $6$\,mT, we first image FGT flakes at $5$\,K, immediately after bringing them below the Curie temperature with a null external field (zero-field cooled). Upon zero-field cooling, ferromagnetic materials tends to minimise the free energy by forming disorganised domain structures which lead to a net-zero stray field. Our stray-field image [Fig.~\ref{fig2}(b)] is consistent with this prediction as we observe regions of net-zero stray field at the interior of the flakes where the domains are below the spatial resolution ($\approx700$\,nm \cite{Lillie2020}). Larger domain features which are above the spatial resolution are located along the edges. 

Starting from this virgin state, we study the magnetization dynamics by taking a series of images between the application of incrementally increasing magnetic fields along the magnetic easy axis ($z$ axis) of FGT [Fig.~\ref{fig2}(c)]. The disorganised domain structure is removed as the strength of the applied field is increased and the flake becomes comprised of a single domain aligned parallel to the direction of the applied field. The delayed transition to a fully magnetised structure suggests domain walls are pinned within the flake creating an energy barrier to domain growth. Thicker ($\approx 18$\,nm) sections form a single domain after the application of $0.08$\,T while the thinner sections ($\approx 8$\,nm or 10 layers) require $0.11$\,T. Following this, the maximum field strength is applied ($1$\,T), resulting in minimal change to the magnetization, thus we infer the flake had already been magnetised at this stage. These quantitative magnetic field maps can be used to reconstruct the out-of-plane magnetization map ($M_z$) using a Fourier inversion method \cite{Thiel2019,BroadwayAM2020,Broadway2020}. For the fully magnetised case in Fig.~\ref{fig2}(c), we find an areal magnetization density ranging between 20 and $50$\,$\mu_{\rm B}$\,nm$^{-2}$, which is over an order of magnitude less than expected from the known magnetization of bulk FGT (see SI, Sec. VI); the reason for this weak magnetization in our thin flake samples is currently not understood. 
With a fully magnetised flake, we similarly take a series of images between the application of fields with increasing strength in opposition to the magnetization to try and reverse the polarity of the flake [Fig.~\ref{fig2}(d)]. Reversal occurs after applying $-0.4$\,T reaffirming previously reported hard ferromagnetic properties \cite{Tan2018}, with no visible thickness dependence.

Again starting from a fully magnetised state, a series of magnetic field images can be taken at increasingly higher temperatures to find the critical temperature at which ferromagnetic order is lost [Fig.~\ref{fig2}(e)]. Here we once again observe a separate behaviour in different sections of the flake; the thinner region becomes demagnetised at a lower temperature ($245$\,K) than the rest of the flake ($265$\,K). Note, our system overestimates the flake's temperature due to separation between the heating element/readout and the FGT sample \cite{BroadwayAM2020}.

Being able to reliably control domain nucleation in precise locations within the flake is a necessary part of studying CIDM. However, the formation of domains under the application of magnetic fields and temperatures to magnetised and demagnetised flakes of FGT [Fig.~\ref{fig2}(c-e)] leads to disorganised structures which is unreliable when trying to repeatably force domains in the same location for repetitive studies of the effect of current injection. By visualising the domain structure we can see Fig.~\ref{fig2}(b) suggests domain pinning may be facilitated by strain within the FGT structure; indeed a correlation can be observed between the domain structure (following zero-field cooling) and the border of flakes located underneath the FGT (WTe$_2$ flakes in this case, outlined in red in Fig.~\ref{fig2}(a)), resulting from external pressure at points of overlap. Thus, local topography features may be used as a way to nucleate domain walls in a reproducible fashion, as required for CIDM studies. We note the WTe$_2$ layer has no observable impact on the magnetic behaviour of FGT studied in Fig.~\ref{fig2}(c-e) otherwise. 

\begin{figure*}[tb]
\centering
\includegraphics[width=0.9\textwidth]{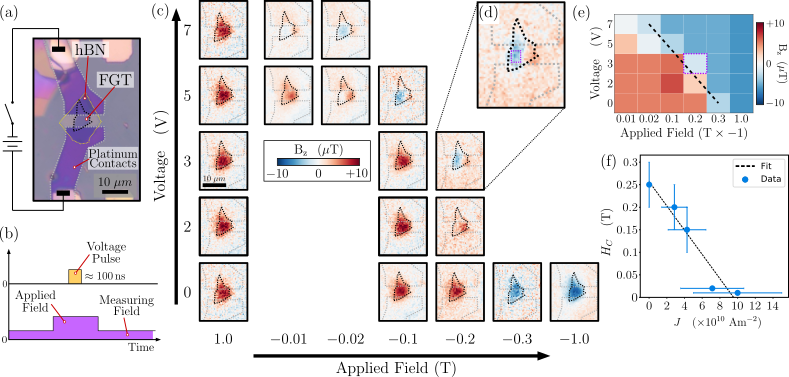}
\caption{{\bf Imaging of current-reduced coercivity in FGT}. (a) Spintronic device fashioned from a flake of FGT ($\approx 8$\,nm or 10 atomic layers), capped with a protective hexagonal boron nitride (hBN) layer, resting on platinum contacts which connect to an external power source. The diamond sensor (not shown here) was placed on top like in Fig.~\ref{fig1}(d). (b) To manipulate the coercive field, short voltage pulses are injected into the FGT while a strong variable field is applied along the easy axis ($z$ axis). The field is then reduced to $B_{\rm NV}^0 = 6$\,mT for imaging (here the NV axis coincides with the $z$ axis). (c) Series of magnetic stray-field images taken with voltage pulses of increasing strength injected under various external magnetic fields. As the applied voltage is increased the field for switching the magnetization decreases. (d) Magnified stray-field image showing the reduced coercivity effect is localised to a particular region of the flake (purple) between the two electrical contacts (grey). (e) Phase diagram of the stray field measured at the sampled location highlighted in d), indicative of domain polarity, as a function of the applied voltage pulse and the applied magnetic field. Dotted line is a guide to the eye representing polarity switching. (f) Coercive field ($H_{\text{c}}$) estimated from the phase diagram in e) plotted against the estimated current density ($J$). A linear fit (dotted line) allows for estimating the spin torque efficiency. Error bars in the coercive field and current density correspond to the step size in the applied field and the uncertainty in estimating the cross-sectional area, respectively.}
\label{fig3}
\end{figure*}

\subsection*{Observation of current-reduced coercivity}

To first demonstrate imaging spintronic effects, we consider current-reduced coercivity, a phenomenon previously observed in FGT and reported to arise from bulk SOTs generated by the injection of currents directly in the FGT flake~\cite{Kaixuan2021, KaixuanAFM2021}. SOTs are supported in FGT by the broken inversion symmetry within its crystal structure \cite{Johansen2019,Martin2021}. We study this effect by constructing a device with a flake of FGT placed between two electrical contacts [Fig.~\ref{fig3}(a)] and taking an image series in a similar fashion to measuring the switching field in Fig.~\ref{fig2}(d), at 5\,K. We first start by magnetizing the flake with a strong field ($1$\,T) along the $z$-axis. Smaller fields are then applied in the opposite direction and a current is injected by applying a short ($100$\,ns) voltage pulse across the two contacts [Fig.~\ref{fig3}(b)]. The field is incrementally increased until the magnetization is reversed. We repeat this process for different voltage pulses. During image collection the field strength is reduced to $6$\,mT.

The image series depicts how as the strength of the voltage pulse is increased, the field required for switching the magnetization is reduced [Fig.~\ref{fig3}(c)]. For no applied voltage ($0$\,V), switching occurs across the entire flake at approximately the same field expected from our previous switching field measurement ($-0.3$\,T here). When applying a $2$\,V pulse, we can now see a reduction in the stray-field emanating from a portion of the flake at $-0.2$\,T. Under the same field, if a $3$\,V pulse is applied, the stray field in the same portion fully reversed. Further increasing the voltage (up to $7$\,V) reduces the field required for switching and increases the area affected.  
Note with a $2$\,V pulse, the magnetization is not fully reversed as the magnitude of the switching field is close to the current-free switching field. Thus we confirm increasing the voltage leads to a decrease in the coercive field.

Interestingly, wide-field NV microscopy enables the observation of current-assisted switching occurring in only a portion of the flake, which we show more clearly with an enlarged image in Fig.~\ref{fig3}(d). In fact the areas where switching does not occur only switch when the $0$\,V switching field of $-0.3$\,T is applied [see SI, Sec. VII]. The localisation of the reduced coercivity effect suggests the reduction occurs in an area of sufficient current density through the FGT. Considering the portion of the flake visibly affected by the current, we can construct a phase diagram by plotting the measured stray field emanating from the flake as a function of the applied field and magnitude of the voltage pulse [Fig.~\ref{fig3}(e)],
highlighting the dependence of the coercive field ($H_c$) on the magnitude of the applied voltage (black dotted line). To facilitate analysis, $H_c$ is plotted against the current density, determined from low-current measurements of the resistance and an estimated cross-sectional area of $\approx 0.1$\,$\mu$m$^{2}$, and fit with a linear relationship [Fig.~\ref{fig3}(f)]. For a maximum injected current density of $10^{10}$\,A$\,$m$^{-2}$ ($7$\,V) we observe a $96\%$ reduction in the coercive field which is within the range of the $50-100\%$ reduction at similar current densities reported in Ref.~\cite{Kaixuan2021}.

The reduced coercivity effect is understood to arise from both Joule heating and SOT \cite{Kaixuan2021}. Joule heating raises the temperature of the FGT, bringing it closer to the ferromagnetic phase transition which softens the characteristically hard magnetic behaviour. Simplistically, the Joule heating effect is expected to increase in strength while raising the magnitude of the voltage pulse as further heating is induced. SOT lowers the energy barrier for magnetization reversal with an effective field $H_{\text{SOT}}$, generated by spin-orbit coupling between the injected electrons and the orbital structure of FGT. The effective field acts on the native magnetic moments and destabilises the magnetization, subsequently reducing the field required for switching \cite{Kaixuan2021}. We can approximate the contribution of the SOT to the reduced coercivity effect in the low current regime using the relationship $H_{\text{c}}(0) - H_{\text{c}}(J) \approx H_{\text{SOT}}$ \cite{Khang2018}. We find a slope of $H_{\text{SOT}}\approx30$\,mT per $10^9$\,A$\,$m$^{-2}$ which to an order of magnitude is consistent with previous measurements \cite{Kaixuan2021}. Note, this is an overestimation of $H_{\text{SOT}}$ as we do not thoroughly account for Joule heating, which could be responsible for up to $60\%$ of the effect in this sample at $J = 10^{10}$\,A\,m$^{-2}$ \cite{Kaixuan2021}. 

\begin{figure*}[tb]
\includegraphics[width=0.9\textwidth]{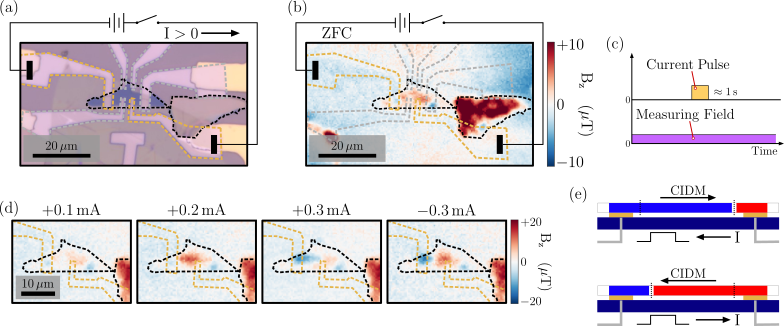}
\caption{{\bf Imaging of current-induced domain wall motion in FGT}. (a) Spintronic device composed of multiple platinum electrical contacts with a flake of FGT ($\approx 12$\,nm or 15 atomic layers) positioned on top. We connect our power supply to only two of the contacts (highlighted in yellow). This is the same device as shown in Fig.~\ref{fig1}(c-f) (images were rotated), with the diamond sensor mounting shown in Fig.~\ref{fig1}(d). (b) Corresponding magnetic image of the zero-field-cooled state prior to any current injection. (c) Current pulses are injected while a small measuring field ($B_{\rm NV}^0 = 6$\,mT applied along the $z$ axis) is being applied. (d) Magnetic images are taken after currents of increasing magnitude are injected. $+0.2$\,mA results in the growth of a domain between the two contacts. Subsequent injection of $+0.3$\,mA results in growth of a new domain of opposite polarity which then shrinks after injecting a current in the opposite direction. (e) A schematic representation of current-induced domain wall motion (CIDM) describing the events resulting in the images taken after the injection of $\pm 0.3$\,mA respectively. The domain walls move in opposition to conventional current.}
\label{fig4}
\end{figure*}

\subsection*{Observation of current-induced domain wall motion}

While imaging the current-reduced coercivity, we also tested whether CIDM could be produced in this device, i.e.\,whether the domain structure can be altered by a current in the absence of magnetic field. However no effect resembling CIDM was able to be observed. We believe this to be due to a sub-optimal device structure as the comparatively large contacts on a small flake of FGT make the nucleation of appropriately sized domains more difficult and decreases the length scale over which we can observe domain-wall motion. Thus, we next studied a spintronic device constructed using the same methods as the device in Fig.~\ref{fig3}(a), but now with multiple, smaller electrical contacts [Fig.~\ref{fig4}(a)]. Prior to current injection, we image the zero-field-cooled state to use as a reference for proceeding images, with barely visible domains close to our spatial resolution [Fig.~\ref{fig4}(b)]. Currents are supplied to the flake of FGT by connecting two of the contacts (highlighted in yellow) to an external power source, and are injected while the small ($6$\,mT) bias field is applied along the $z$-axis [Fig.~\ref{fig4}(c)]. After current injection the magnetic stray field is imaged to determine whether it has resulted in a change in the domain structure. Taking a series of images after injecting currents of both increasing magnitude and in alternating directions [Fig.~\ref{fig4}(d)], we observe the growth of domains between the two contacts with polarity both aligned and in opposition to the constant bias field. 

Namely, starting from the zero-field-cooled state, a $+0.1$\,mA current is briefly ($\sim1$\,s) applied to the unmagnetised flake. The resulting micromagnetic structure shows little deviation from the zero-field-cooled state suggesting the supplied current is insufficient to overcome the energy barrier to drive domain wall motion \cite{Kim2013}. At this stage, small domains of positive (red, i.e.\,pointing towards $+z$) and negative (blue, $-z$) polarity are visible near the left and right contacts, respectively, with a large white domain in between, indicating there are at least 3 domain walls between the contacts. Increasing the current to $+0.2$\,mA leads to the growth of a singular domain (red, $+z$ polarity) between the contacts. Reversing the direction of the current at this value leads to no significant change in the domain structure [see full image series in SI, Sec. VII]. Further increasing the current to $+0.3$\,mA leads to the growth of a new domain with the opposite polarity (blue, $-z$) to the previous domain. By changing the direction of the current the domain is shrunk by growing an adjacent opposing ($+z$) domain. The growing (shrinking) of domains after the injection of currents suggests the current is pushing a domain wall through the flake of FGT.  
As the bias field is present during the current injection (pointing towards $+z$), we can rule out trivial current-reduced coercivity or thermal effects, both of which would exclusively lead to domain growth with positive polarity, in alignment with the bias field. We also note domains external to the two electrical contacts remain unchanged throughout. Considering the injection of $\pm0.3$\,mA, we suggest the positive current has pushed a domain wall from left to right before the negative current then pushes it back (right to left) [Fig.~\ref{fig4}(e)]. Consequently, the domain wall is determined to be moving against the direction of conventional current.

CIDM in metallic nano-wires and multilayer structures is generally attributed to two driving mechanisms, STT and SOT \cite{MironNM2011,Emori2013}. In each case, a spin current is generated by polarising the magnetic moments of the injected electrons. For STT, the current is polarised when passing through the magnetic material, a torque is then generated by an effective magnetic field which acts upon the domain wall. In the case of SOTs, spin polarisation arises from spin-orbit coupling, generally in an adjacent non-magnetic layer. Similar to STT, the spin current generates an effective field which creates a torque at the domain wall. Both torques act on the local magnetic moment within the domain wall forcing it to move in a direction dependent on the polarity of the spin current and the domain wall geometry \cite{Manchon2019}. To characterise the spin torque efficiency (regardless of its type and origin) in CIDM experiments, it is common to consider the equivalent magnetic field needed to be applied to move the domain wall. Here, the depinning field is at least 6 mT, the value of the bias field. If we assume our device has a cross-sectional area of $\approx 0.1$\,$\mu$m$^{2}$ and consider the case where $0.3$\,mA is applied to grow a domain opposed to the bias field, i.e.\,the current needs to be equivalent to a $-12$\,mT applied field at least, we find a spin-torque efficiency of $4$\,mT per $10^9$\,A\,m$^{-2}$. This is nearly two orders of magnitude more efficient than in conventional multilayer systems with perpendicular magnetic anisotropy such as Pt/Co/AlO$_x$ and Ta/CoFe/MgO \cite{MironNM2011,Emori2013} and is similarly also more efficient than in synthetic antiferromagnetic systems \cite{YangNN2015}. In these systems, SOT arising from the spin Hall effect in the heavy metal underlayer is believed to be the dominant mechanism \cite{Emori2013,Manchon2019}.

The reason for the high spin-torque efficiency observed in our experiments may be due to the bulk SOT generated by the spin Hall effect, in contrast to the interfacial SOT at play in multilayer systems. The existence of bulk SOT in FGT was recently confirmed by transport measurements using harmonic analysis, with effective fields of up to $0.5$\,mT per $10^9$\,A\,m$^{-2}$ reported \cite{Martin2021}, which compares favourably to our value determined above. However, the existence of a spin torque does not guarantee domain wall motion \cite{Manchon2019}, and the exact mechanism driving CIDM in our experiments remains unclear.  
% SOT is typically a more efficient driver for CIDM than STT \cite{Emori2013}. For our measurement in FGT, if we assume a cross-sectional area of $\approx 0.1$\,$\mu$m$^{2}$ the minimum current density at which we observe CIDM is $\approx 2$\,mA\,$\mu$m$^{-2}$, which is $1000$ times less than the current density used for SOT-driven domain wall motion in magnetic multilayers \cite{MironNM2011, Zhaochu2020, Guan2021}.
%The improved efficiency can at least in part be attributed to the low dimensionality of the FGT increasing the susceptibility to current induced effects, however it is insufficient reasoning to determine if SOT is the responsible mechanism.
Using symmetry arguments based on the $D_{3h}$ point group of monolayer FGT \cite{Laref2020,Ado2021} [see SI, Sec. X], it is found the chiral spin spirals typically associated with SOT-driven domain-wall motion are not stabilized, suggesting there may be additional unexplored mechanisms contributing to the observed behaviour. Possible mechanisms responsible for CIDM in FGT may involve interfacial effects, strain, material defects or interlayer interactions (within FGT). Even in ideal FGT, current-induced torques is still a subject of intense research \cite{Martin2021} as illustrated by the recent description of the so-called orbital torque \cite{Saunderson2022}. A study of the effect of these current-induced torques on domains walls in ideal and real samples is beyond the scope of the present work. Thermal effects may also play a nontrivial conflating role. 

Further studies of CIDM in this particular device were prevented after excessive Joule heating caused the device to fail [see SI, Sec. IX], when raising the current above $0.3$\,mA. Subsequent FGT devices we have tested have shown reproducibility of the CIDM effect is in part dependent on device structure, including: electrical contact structure, electrical contact placement, and flake size. For future studies the current-reduced coercivity behaviour could be exploited to reliably nucleate domains at precise locations. Such amendments will enable more thorough and systematic studies of CIDM with widefield NV microscopy, which will help inform theoretical studies of the underlying mechanisms.

\section{Conclusions}

In this work, we resolved the entire micromagnetic structure of thin flakes of FGT using widefield NV microscopy to enable the study of spintronic phenomena, namely current-reduced coercivity and current-induced domain wall motion. We have respectively confirmed and demonstrated how the existence of efficient current-induced torques enables low power control of the magnetic domains, a desirable feature for future 2D vdW magnet devices. However, the preliminary study of the ferromagnetic properties demonstrate potential challenges for future studies and device implementation. The working temperature of FGT based devices has to be below room temperature to facilitate the critical behaviour of ferromagnetism and, reliable nucleation of domains within a flake of material requires an external force such as strain. Ionic gating and integration with topological insulators have been shown to raise the Curie temperature of FGT \cite{Yujun2018,Haiyu2020} as has increasing the iron content \cite{Nair2021}. Controlling the nucleation of domains is likely better facilitated by the current-reduced coercivity phenomenon which we have shown to only affect areas of the flake with significant current flow. The bulk SOT native to FGT, which is responsible for the current-reduced coercivity, along with other types of current-induced torques, require further investigation to determine whether they may play a role in CIDM and if it may be able to drive other phenomena such as zero-field switching \cite{Liang2021}. While the widefield NV microscope is useful for observing changes in the micromagnetic structure providing real-space insights, complementing NV measurements with techniques that directly probe current-induced torques such as ferromagnetic resonance and harmonic analysis in Hall effect measurements \cite{Manchon2019} will be necessary to obtain a fuller picture.

%\section*{Data availability statement}

%The data that support the findings of this study are available upon reasonable request from the authors.

\begin{acknowledgments}
This work was supported by the Australian Research Council (ARC) through grants CE170100012, CE170100039, and FT200100073. The work was performed in part at the RMIT Micro Nano Research Facility (MNRF) in the Victorian Node of the Australian National Fabrication Facility (ANFF) and the RMIT Microscopy and Microanalysis Facility (RMMF).  I.O.R. and A.J.H. are supported by an Australian Government Research Training Program Scholarship. S.C.S gratefully acknowledges the support of an Ernst and Grace Matthaei scholarship.
\end{acknowledgments}

\bibliography{references_editable}

\clearpage
\onecolumngrid

\begin{center}

\textbf{\large Supplementary Information for the manuscript ``Imaging current control of magnetization in Fe$_3$GeTe$_2$ with a widefield nitrogen-vacancy microscope''}

\end{center}
%%%%%%%%%% Merge with supplemental materials %%%%%%%%%%
%%%%%%%%%% Prefix a "S" to all equations, figures, tables and reset the counter %%%%%%%%%%
\setcounter{equation}{0}
\setcounter{section}{0}
\setcounter{figure}{0}
\setcounter{table}{0}
\setcounter{page}{1}
\makeatletter
\renewcommand{\theequation}{S\arabic{equation}}
\renewcommand{\thefigure}{S\arabic{figure}}

\section{Diamond samples}

The NV-diamond substrates used in this work were made from type-Ib, single-crystal diamond plates grown by high-pressure, high-temperature synthesis, with $\{111\}$-oriented polished faces (polishing done by Technical Diamond Polishing, UK) for main text figures 1, 3 and 4, or $\{100\}$-oriented polished faces (polishing done by Delaware Diamond Knives, USA) for main text figure 2. The thickness of the polished plates was about $100\,\mu$m. The near-surface NV layer was created by irradiating the plates with 2 MeV antimony ions (main text figures 1, 3 and 4) or 100 keV carbon ions (main text figure 2) following the same parameters as described in Refs. \cite{BroadwayAM2020} and \cite{Abrahams2021}, respectively. This gives NV layers about 200 nm and 500 nm thick, respectively. Following irradiation the diamonds were annealed, laser cut into (laterally) smaller plates, and  acid cleaned. The $\{100\}$-oriented polished face diamond used for data collection in main text figure 2 has a $150$\,nm Al/Al$_2$O$_3$ patterned coating on the surface as described in Ref.~\cite{BroadwayAM2020}, which separates the diamond and the sample during measurement. The purpose of this structure is to provide reference coordinates to easily locate the sample, as well as to block the laser from directly interacting with the sample, which reduces laser-induced heating and also helps to decouple any measurement artefacts from potential changes in the optical response of the sample during the experiments (e.g. following current injection).

\section{FGT samples and devices}
\label{SI: FGT prep}

All FGT, hBN, and WTe$_2$ microflakes were mechanically exfoliated in an Argon-filled glove box with oxygen and water levels below 0.1 parts per million. The details of the bulk crystals used for exfoliation can be found in Refs. \cite{Tan2018,Tan2021}. For the non-contacted samples (main text figure 2), the FGT and WTe$_2$ flakes were picked up in sequence by PC (polycarbonate) films mounted on PDMS (polydimethylsiloxane) stamps and then released onto the diamond substrate coated with a Al$_2$O$_3$/Al grid as in Ref. \cite{BroadwayAM2020}. The PC film was left on the diamond for the NV measurements, acting as a protecting layer to the FGT.

For the spintronic devices (main text figures~1,3, and 4), the bottom Pt contacts ($\approx 10$\,nm thick) were prepared on SiO$_2$/Si substrates by standard e-beam lithography and sputtering techniques. When assembling the spintronic devices, we used PC/PDMS to align and pick up the hBN and FGT layers in sequence, the PC/hBN/FGT heterostructure was then transferred onto the Pt electrodes. The PC film was dissolved, and a layer of PMMA (Polymethyl methacrylate) A7 was spin-coated onto the surface for further protection of the devices. The diamond is then secured to the device as described in Sec. \ref{si:diamond}.

\section{Diamond interfacing}
\label{si:diamond}

\begin{figure*}[t!]
\includegraphics[width=0.9\textwidth]{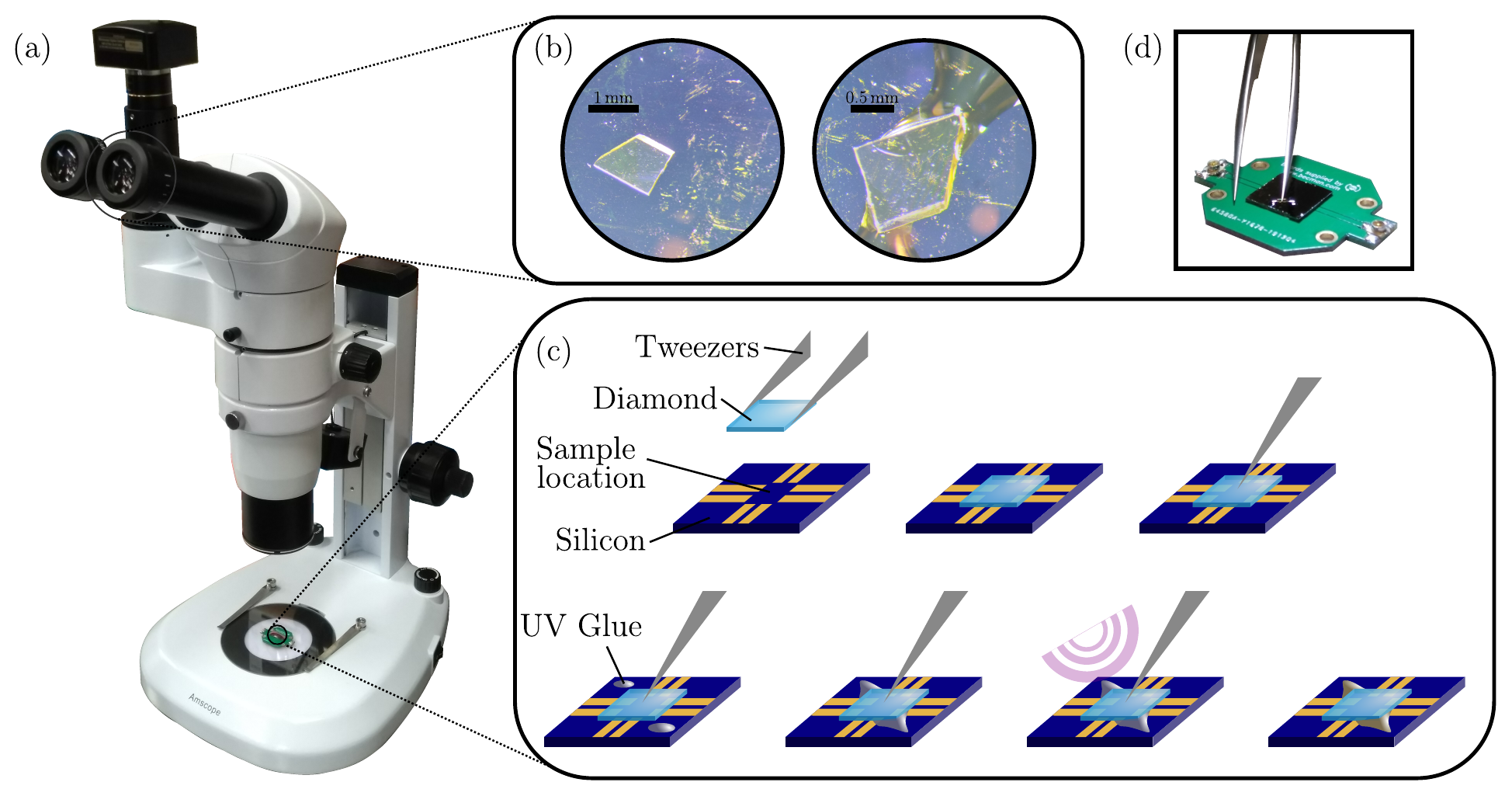}
\caption{{\bf Diamond interfacing for spintronic devices.} (a) A microscope is used for mounting the diamonds on the spintronic devices due to the small size of the diamonds and the necessary precision of placement. (b) Optical micrographs of a diamond before and after gluing on a silicon substrate. The glue holds the diamond in place and helps to ensure a minimal separation between the NV layer and sample. (c) Step-by-step method for precisely and reliably gluing diamonds onto the surface of a sample. The diamond is placed over the desired area using a pair of tweezers and then held in place by a second pair of tweezers. UV setting glue (Thorlabs NOA63) is then deposited near the corners of the diamond and then dragged to the diamond until it just contacts the edges. The glue is then set using UV irradiation securing the diamond in place. (d) Optical image of tweezers holding a diamond in place during the gluing process.}
\label{SI_fig1}
\end{figure*}

Studying vdW magnets previously has employed a diamond-sample interfacing method where the magnetic material is transferred onto the surface of the diamond \cite{BroadwayAM2020} (as we also did in main text figure 2) which ensures a fixed standoff distance between the sample and NV layer optimised for maximum sensitivity and spatial resolution \cite{Scholten2021}. Due to the more complex nature of the spintronic devices, this method is less feasible as it would require time consuming and complex fabrication techniques. Instead, here the spintronic devices are fabricated on a standard SiO$_2$/Si substrate and the diamond is dropped onto the surface of the device.  A fixation method is then used to ensure a minimal standoff and to prevent the diamond from being removed from the surface when the device is being loaded into the cryogenic chamber. For this purpose small diamonds ($\approx 1$\,mm$^2$) are used to reduce the probability of surface unevenness creating additional unwanted standoff separation. Thus due to the small areal size of the diamond and the precision with which the diamond must be placed, the procedure is carried out under a microscope [Fig.~\ref{SI_fig1}(a)]. Once the diamond is at the desired location it is glued at the corners [Fig.~\ref{SI_fig1}(b)] using the method pictorially depicted in Fig.~\ref{SI_fig1}(c,d). The presented method is a simple, repeatable way of interfacing a diamond with a sample without implementing `lucky drop' based methods. Images taken in main text figures~1,3, and 4 were taken by interfacing a diamond in this way.

\section{Experimental setup}

All NV measurements were carried out on a cryogenic widefield microscope, described in detail in Ref. \cite{Lillie2020}. The closed-cycle cryostat (Attocube attoDRY1000) has a base temperature of about $5$\,K during NV measurements. Magnetic fields of up to 1~T in any direction can be applied using a superconducting vector magnet. A 532~nm laser (Laser Quantum Ventus) allows NV initialisation and readout and pulse control is provided by a fibre-coupled acousto-optic modulator (AAOpto MQ180-G9-Fio). The laser is passed through a beam scrambler (Optotune LSR-3010) to remove optical interference patterns that reduce the illumination uniformity, before being focussed using a low-temperature microscope objective (Attocube LT-APO/VISIR/0.82) to obtain roughly even laser illumination over a region covering most of the $\sim$100~$\mu$m field of view. The laser power density at the NV layer, following losses along the optical path, is estimated at around 1~kW/cm$^2$. NV PL is collected through the same objective, filtered (709/167~nm) and imaged onto a sCMOS camera (Andor Zyla 5.5-W USB3). 

For main text figure 2, the diamond substrate (supporting the FGT samples) is glued onto a glass coverslip patterned with a gold microwave resonator to facilitate NV spin state driving, which is then connected to a printed circuit board. For main text figures 1, 3 and 4, the microwave is delivered via a coplanar waveguide built in the circuit board. A signal generator (Rohde \& Schwarz SMB100A) provides the microwave signal, gated by a switch (Mini-Circuits ZASWA-2-50DR+) and amplified (Mini-Circuits HPA-50W-63). Pulse sequences (including synchronising with camera acquisition) were programmed onto and controlled by a SpinCore PulseBlaster ESR-PRO 500 MHz card.

\section{NV measurement details}

All magnetic measurements presented were obtained using pulsed optically detected magnetic resonance (ODMR). A 20~$\mu$s laser pulse (chosen to achieve a balance between NV initialisation and readout contrast) was followed by a microwave $\pi$ pulse ($\approx 300$~ns) of a given frequency. This sequence was then repeated to fill the 30~ms camera exposure. A second camera exposure follows with a no-microwave sequence to act as a reference, against which the first exposure can be normalised. This was then repeated over a desired range of frequencies to build up an ODMR spectrum. Following sufficient signal acquisition (typically several hours; thousands to tens of thousands total sweeps), a magnetic image can be obtained by extracting the ODMR spectrum at each camera pixel and fitting the resonance frequencies. 

All measurements were taken under a low bias field (strength $\approx 6$~mT) along a chosen NV axis, i.e. one of the $\langle 111\rangle$ directions of the diamond crystal. For the $\{100\}$-oriented surface (as in main text figure 2), the chosen NV axis points at 54.7$\degree$ from the $z$ axis, and  45$\degree$ from the $x$ axis. For the $\{111\}$-oriented surface (as in main text figures 1, 3 and 4), we choose the NV axis that is perpendicular to the surface, i.e. the $z$ axis. For this low bias field, we can address both the $\ket{0}\rightarrow\ket{-1}$ and $\ket{0}\rightarrow\ket{+1}$ NV ground state spin transitions, producing two resonances at frequencies $f_-=D-\gamma_{\rm NV}B_{\rm NV}$ and $f_+=D+\gamma_{\rm NV}B_{\rm NV}$, where $D$ is the zero-field splitting, $\gamma_{\rm NV} = 28.033(3)$~GHz/T is the NV gyromagnetic ratio (approximately equal to the free electron value, $\gamma_e$), and $B_{\rm NV}$ is the total magnetic field projection along the NV axis~\cite{Scholten2021}. These resonances were fit by Lorentzian functions with bounded frequencies, amplitudes, and widths. Taking the difference of the two frequency maps gives the desired $B_{\rm NV}=(f_+ - f_-)/2\gamma_{\rm NV}$ map. In practice we subtract the bias field and any background variation (e.g. due to variations in laser intensity or microwave driving artificially altering the fit frequencies) across the field of view to obtain the signal only due to FGT flakes. From the ODMR spectra a map of $D=(f_+ + f_-)/2$ can also be obtained, which contains information regarding variations in the crystal strain within the diamond lattice, which is sometimes intrinsic to the crystal but more commonly in our samples is due to the Al grid deposition. These features are typically large in magnitude compared to the magnetic signal from target flakes. The ODMR data is fit with a Lorentzian function with an added linear component which helps to remove image artifacts associated with fitting.  

% Additional background subtraction methods may be applied which directly consider the geometry of the flake to emphasise the micromagnetic structure. These isolate the magnetic features of interest and interpolate the magnetic background which is then subtracted from the entire image. 

\section{Reconstruction of magnetization}

\begin{figure*}[h!]
\includegraphics[width=0.6\textwidth]{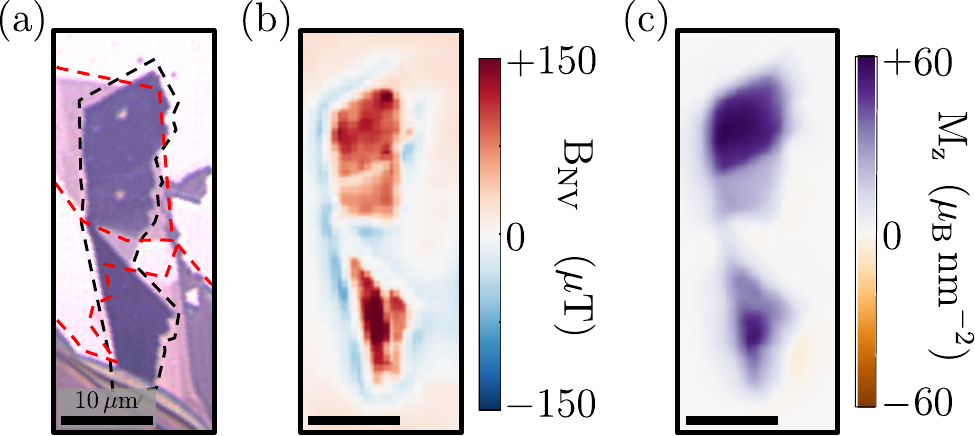}
\caption{{\bf Magnetization reconstruction.} (a) Optical image of the flake of FGT (black outline) presented in main text figure~2. (b) The corresponding stray-field image of the FGT flake after applying a $1$\,T magnetic field along the axis of perpendicular anisotropy ($z$-axis). (c) Reconstructed image of the $z$ component of the magnetization from the stray-field image in (b).}
\label{SI_mag_recon}
\end{figure*}

The flake of FGT imaged in main text figure~2 [reproduced in Fig.~\ref{SI_mag_recon}(a)] was fully magnetized by application of a $1$\,T field in the positive $z$ direction [Fig.~\ref{SI_mag_recon}(b)]. Using a Fourier inversion method we can reconstruct the source magnetization from the stray field measured at the NV layer \cite{Broadway2020}. Reconstructing the $z$ component of the magnetization [Fig.~\ref{SI_mag_recon}(c)], clearly shows a spatial dependence of the strength of the magnetization which we attribute to areas of different thickness/number of layers. For instance, in the top part of the image the thick (thin) region of the flake with 23 (10) layers is found to have an average areal magnetization density of $50$\,$\mu_{\rm B}$\,nm$^{-2}$ ($20$\,$\mu_{\rm B}$\,nm$^{-2}$). From earlier measurements of FGT we can assume each iron ion should actually contribute $1.5$\,$\mu_{\rm B}$ on average (as in bulk FGT at low temperatures \cite{Chen2013}), thus from the lattice parameters of FGT ($a=b=0.3991$\,nm and $c=1.633$\,nm) and the density ($7.3$\,Mgm$^{-3}$) \cite{Villars2016}, we expect an areal magnetization density of $33$\,$\mu_{\rm B}$\,nm$^{-2}$ per monolayer. This would give $760$\,$\mu_{\rm B}$\,nm$^{-2}$ ($330$\,$\mu_{\rm B}$\,nm$^{-2}$) for a saturated 23-layer (10-layer) flake, about 15 times larger than we measured. A possible explanation for this discrepancy is partial sample degradation but the absence of thickness dependence in this reduction, and the absence of significant reduction of the Curie temperature compared to bulk FGT make this explanation unlikely. A measurement/analysis error is also possible (sources of error in NV imaging are discussed in Ref. \cite{Scholten2021}) but it is unlikely to alone explain a 15-fold apparent reduction. We speculate that there may exist a partial compensation of magnetization between the different layers composing the flakes at low field, e.g. due to an antiferromagnetic interlayer coupling; such a coupling has been observed in partially oxidised FGT samples \cite{Kim2019b} but further work is required to confirm if it could be present in our samples. 

\section{Current-reduced coercivity}

\begin{figure*}[h!]
\includegraphics[width=0.9\textwidth]{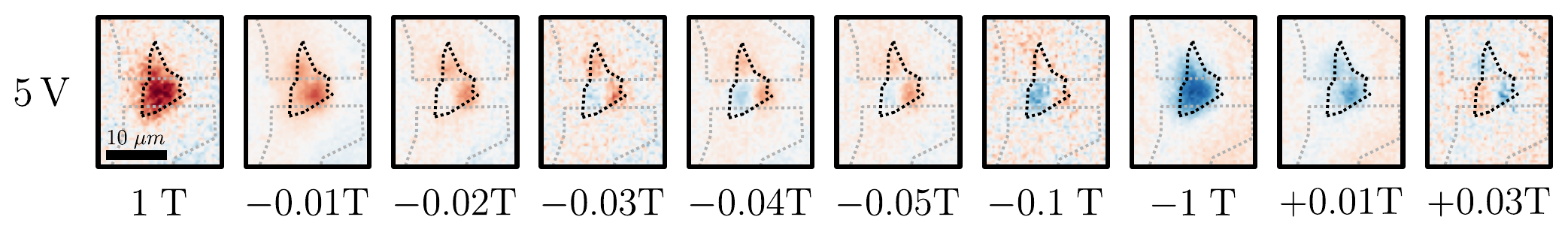}
\caption{A more complete series of the current-reduced coercivity effect under a $5$\,V pulse including switching in the opposite direction after the application of positive fields. The conditions are as described in main text figure 3.}
\label{SI_fig2}
\end{figure*}

In main text figure 3, we studied the switching from positive to negative polarity. To demonstrate the observed hysteresis behaviour of the current-reduced coercivity effect as in Ref.~\cite{Kaixuan2021}, here we show a more complete series of images used in part in main text figure~3 [Fig.~\ref{SI_fig2}], including an example of magnetization switching from the negatively magnetized state. Namely, after performing the switching experiment with the application of $5$\,V pulse starting from the positively magnetized state (red) the flake is negatively magnetized (blue), and positive fields are applied with the concurrent voltage pulse. As shown in the final panel the magnetization in portion of the flake between the contacts is reversed with an applied field of $+0.03$\,T, the same magnitude as expected from the reversed case. 

\section{Current-induced domain-wall motion}
\label{CIDM_SI}

\begin{figure*}[b!]
\includegraphics[width=0.9\textwidth]{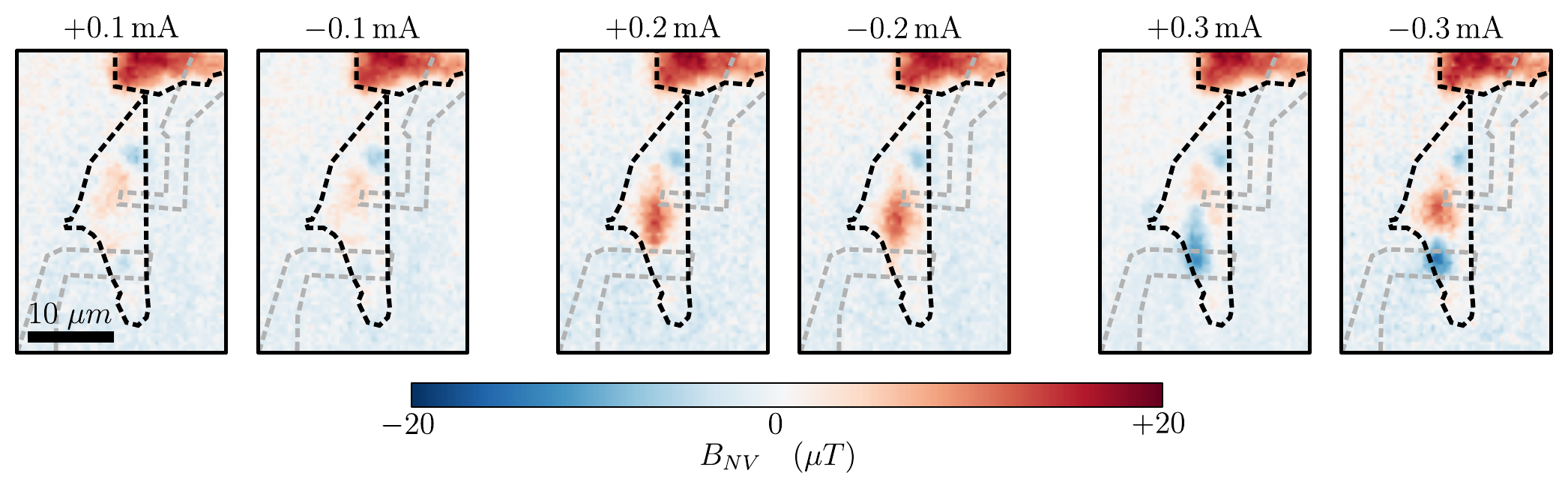}
\caption{Series of magnetic images of CIDM with both the positive and negative currents. The conditions are as described in main text figure 4.}
\label{SI_fig3}
\end{figure*}

Extending main text figure 4, in Fig. \ref{SI_fig3} we show a series of stray-field images showing how injection of smaller currents ($\pm 0.1$\,mA) leaves the magnetic structure indistinguishable from the unmagnetized zero-field cooled state, suggesting the current carried insufficient energy to depin a domain-wall \cite{Kim2013}. CIDM is observed in FGT after the injection of $+0.2$\,mA where there is clear growth of a domain as indicated by the appearance of an area with a stronger positive magnetic field. When the $-0.2$\,mA current is injected, there is no change with respect to the previously injected current. Injecting $+0.3$\,mA creates a notable change in the magnetic structure as the current has caused the growth of a negatively signed domain from the lower contact which is then shrunk by the growth of another domain from the upper contact when $-0.3$\,mA is injected. Note, the other pinned domain is over an unused contact and has remained unchanged throughout all of these measurements, giving clear indication the observed domain-wall motion is a direct result of the injected currents.

To understand the mechanics of the CIDM it is more intuitive to consider the images after the injection of $\pm0.3$\,mA first. Prior to these currents, a large domain was grown between the two contacts with a positive stray-field. Following the injection of $+0.3$\,mA, this domain is seemingly replaced by one with the opposite sign. Given there is no obvious domain at the top contact with this sign and there is one pinned at the bottom contact, this suggests the domain was likely grown from the bottom contact by pushing the domain-wall towards the top contact following the direction of electron flow. Similarly, the injection of $-0.3$\,mA shrinks this domain by pushing the domain wall back towards the bottom contact again following the flow of electrons. Establishing the direction of domain-wall motion relative to the electron flow helps to explain the domain growth at lower currents. As the domain-wall motion follows the electrons, it can be assumed the injection of $+0.2$\,mA grows the domain from the bottom contact at the left edge of the flake, right at the point where the contact initially connects with the flake of FGT, instead of the pinned domain. Therefore the current is likely not flowing through the entirety of the contact and is instead flowing into the flake, following the path of least resistance. When the direction of the current is reversed, there is a small change as the domain grows slightly into the bottom contact which suggests the current is passing back through the same domain following the the same current path as previously described.

\section{Electrical response and Joule heating}
\label{joule_heating_SI}

\begin{figure*}[b!]
\includegraphics[width=0.9\textwidth]{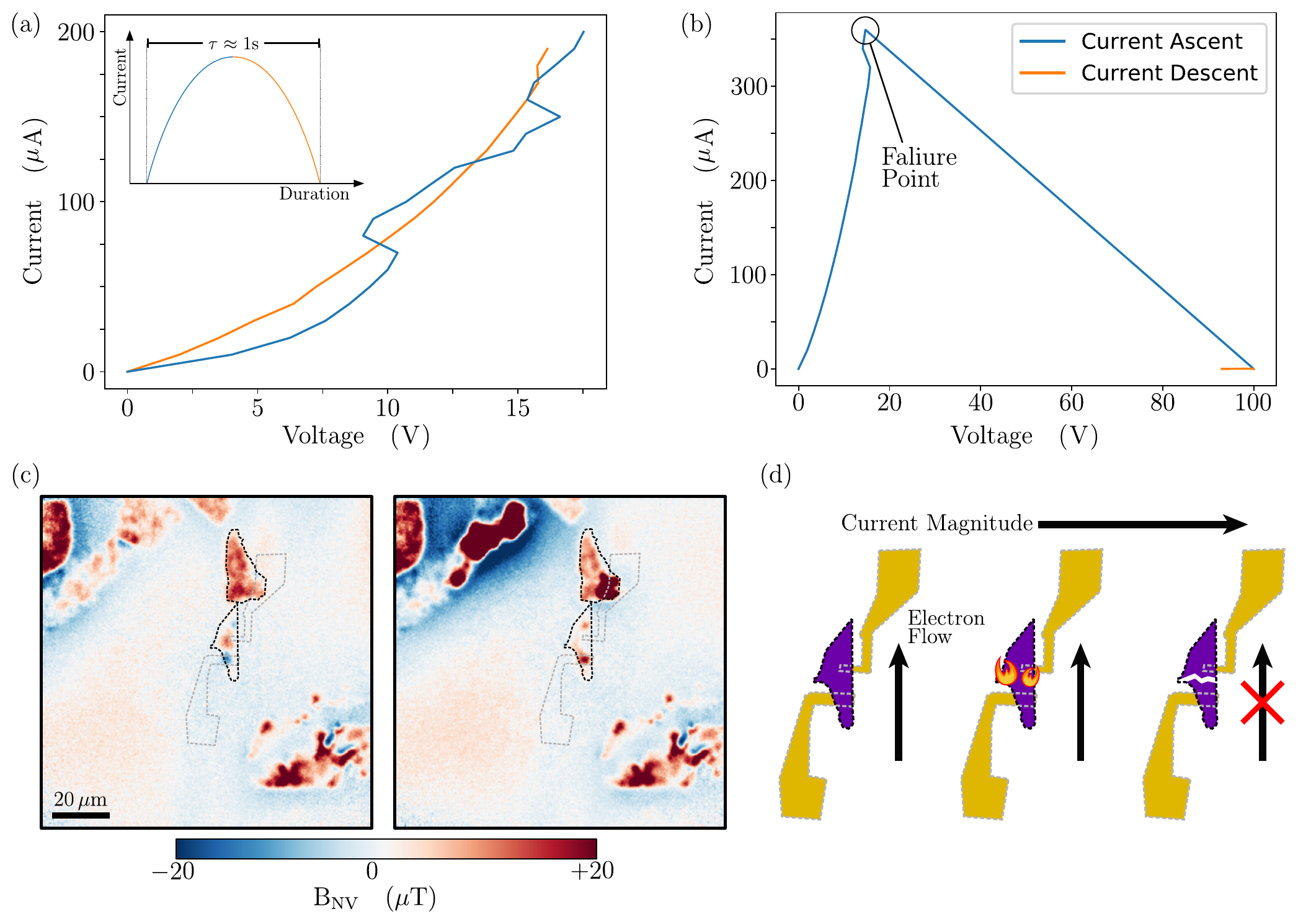}
\caption{{\bf Effects of current injection and Joule heating} (a) IV curves recorded during current sweeps from $0 \rightarrow 200$\,$\mu$A (blue) and from $200 \rightarrow 0$\,$\mu$A (orange). The measurement is recorded continuously over a period of approximately $1$\,s (inset). (b) Attempted IV curve measurement for a current sweep from  $0 \rightarrow 400$\,$\mu$A which led to critical failure of the device creating an open circuit and preventing further current injection. (c) Stray-field images of the device before (left) and after (right) the critical failure event. Comparing the two images shows the event led to alteration of the micromagnetic structure within the flake of FGT. (d) Schematic representation of the proposed critical failure event where Joule heating led to destructive burning of the device.}
\label{SI_fig4}
\end{figure*}

For the CIDM experiment, currents were supplied by  a source-measure unit (SMU) (Keithley 2450 SourceMeter). The SMU supplied currents through a sweep function which applies a series of currents starting from zero up to a set maximum over a short but fixed time interval before returning to zero along the same path. When injecting currents the SMU simultaneously records the associated voltage and generates a current-voltage (IV) plot [Fig.~\ref{SI_fig4}(a)]. The IV curve exhibits nonlinear behaviour on both the rising and falling parts of the current sweep showing the device exhibits non-ohmic behaviour. The jaggedness of the rising part of the curve in comparison to the falling part is indicative of some irreversible, non-repeatable events resulting from the increase in current (annealing process).
%We believe this to be a heating related effect where the organic material on the surface of the device was being burned off. 

As addressed in the main text the devices are fragile and susceptible to destruction. Here we show directly the event which destroyed the device used to observe CIDM in the main text and suggest a possible explanation for what occurred. Upon observing domain wall motion after the injection of $\pm0.3$\,mA the current was raised to $+0.4$\,mA in an attempt to further probe the current dependence. In hindsight this was a mistake as the injection of $+0.4$\,mA broke the device as indicated by the IV curve for this current sweep [Fig.~\ref{SI_fig4}(b)]. The IV curve in Fig.~\ref{SI_fig4}(b) has similar lineshape to the previous IV curve in Fig.~\ref{SI_fig4}(a) up until the current sweep reaches $0.36$\,mA where there was a sudden loss of current as the voltage was increased. Attempting to inject current after this event confirmed the circuit had been broken as no current passed through the device. The non-ohmic behaviour of the device suggests the nonlinear rise in the resistance would also increase the local heating within the device. Thus we presume the device was heated sufficiently to fracture the flake of FGT.

To further confirm overheating of the device destroyed the flake we can compare the stray-field images before and after the event [Fig.~\ref{SI_fig4}(c)]. In the before image [Fig.~\ref{SI_fig4}(c) left] there are clear negative domains (blue) in the central flake of FGT. After attempted the injection of $+0.4$\,mA, these domains are now positive (red); furthermore magnetic features surrounding the flake are more strongly magnetised with positive polarity. Thus it is concluded the injected current caused significant heating, enough to bring the flakes close to or over their Curie temperature, temporarily softening the magnet or putting it in an unmagnetised state. The $6$\,mT measuring field, which was left on during current injection then forced the alignment of magnetic moments, leaving the flakes positively magnetised after they cooled back down. As current no longer passes through the device we additionally believe the heating caused a fracture in the flake of FGT ultimately destroying the device [Fig.~\ref{SI_fig4}(d)].

\section{chiral magnetic exchange in D$_{3h}$ crystals}

A symmetry analysis is performed below in order to try and determine the responsible mechanism for the observed CIDM. The analysis considers an ideal FGT crystal with D$_{3h}$ symmetry and examines the bulk SOT generated by a current in the FGT \cite{Martin2021}. For sustained SOT driven domain wall motion to occur, the analysis should reveal the presence of a stabilised chiral domain wall or spin spiral. However the terms generated in the gradient expansion do not stabilize the usual N{\'e}el/Bloch spirals but are instead canted. This suggests that the observed domain wall motion involves a mechanism yet to be determined and not captured by SOT in ideal FGT. Possible origins include symmetry breaking due to strain or material defects, interfacial effects (between FGT and substrate or top hBN layer), interlayer interactions (within FGT), orbital torques \cite{Saunderson2022}, and thermal effects. These will be the subject of future work. 

\begin{table}[b!]
    \begin{tabular}{c|c|c|c|c|c|c|c|c|c|}
         \hline
         \textbf{D$_{3h}$} & \textbf{E} & \textbf{2C$_{3}$} & \textbf{3$C_{2}$'} & \textbf{S$_{h}$} & \textbf{2S$_{3}$} & \textbf{S$_{v}$} & \textbf{Linear} & \textbf{Quadratic} & \textbf{Cubic} \\
         \hline
         A$_1$' & 1 & 1 & 1 & 1 & 1 & 1 & - & $z^2$, $x^2 + y^2$ & $x(x^2 - 3y^2)$\\
             &   &   &   &   &   &   &   & $m_z^2$, $m_x^2 + m_y^2$ &  \\
        \hline
        A$_2$' & 1 & 1 & -1 & 1 & 1 & -1 & $m_z$ & - & $y(3x^2 - y^2)$\\
             &   &   &   &   &   &   &   &  & $m_z^3$, $m_z(m_x^2 + m_y^2)$ \\
        \hline
        E' & 2 & -1 & 0 & 2 & -1 & 0 & $x$, $y$ & $x^2 - y^2$, $xy$ & $xz^2$, $yz^2$\\
             &   &   &   &   &   &   &   & $m_x^2 - m_y^2$, $m_x m_y$ & $x(x^2 + y^2)$, $y(x^2 + y^2)$ \\
        \hline
        A$_1$'' & 1 & 1 & 1 & -1 & -1 & -1 & - & - & $m_x(m_x^2 - 3m_y^2)$\\
             &   &   &   &   &   &   &   &   &  \\
        \hline
        A$_2$'' & 1 & 1 & -1 & -1 & -1 & 1 & $z$ & - & $z^3$, $z(x^2 + y^2)$ \\
             &   &   &   &   &   &   &   &   & $m_y(3m_x^2 - m_y^2)$ \\
        \hline
        E'' & 2 & -1 & 0 & -2 & 1 & 0 & $m_x$, $m_y$ & $xz$, $yz$ & $zxy$, $z(x^2 -y^2)$ \\
             &   &   &   &   &   &   &   & $m_x m_z$, $m_y m_z$  & $m_x m_z^2$, $m_y m_z^2$ \\
             &   &   &   &   &   &   &   &   & $m_x(m_x^2 + m_y^2)$, $m_y (m_x^2 + m_y^2)$
    \end{tabular}
    \caption{Character Table of $D_{3h}$. $(x,y,z)$ are the components of a polar vector, and $(m_x, m_y, m_z)$ are the components of an axial vector.}
    \label{tab1}
\end{table}

\paragraph*{Symmetry analysis.} A monolayer of Fe$_3$GeTe$_2$ belongs to the $D_{3h}$ point group whose character table is given in Table \ref{tab1}. This table summarizes how irreducible representation of linear, quadratic and cubic functions transform upon symmetry operations of the $D_{3h}$ point group. We note that in multilayer FGT each layer
satisfies the symmetries, and given the interlayer coupling is weak the analysis is expected to remain valid \cite{Kaixuan2021}. In the small gradient limit, the magnetic energy can be expended in gradients of the magnetization on the form 

\begin{equation}
    E_M \sim \sum_{\substack{\alpha,\beta,\gamma \\ n,m,l}} m_\alpha^n \partial_\beta^m m_\gamma^l
\end{equation}

Where $\alpha, \beta,\gamma=x,y,z$ and $n,m,l\in\mathbb{N}$. In the absence of external magnetic field, $E_M$ only involves terms that are even in magnetization components, so that $E_M (\Vec{m})=E_M (-\Vec{m})$. Since the magnetization is an axial vector and the spatial gradient is a polar vector, each term in $E_M (\Vec{m})$ can be expressed as the combination of the irreducible representations of these vectors components that transforms as $A_1$’. By applying this method, we find the combinations shown in table \ref{tab2}. 

\begin{table}[t!]
    \begin{tabular}{c|c|c|c|c|c|c|c|c|}
         \hline
         \textbf{D$_{3h}$} & \textbf{E} & \textbf{2C$_{3}$} & \textbf{3$C_{2}$'} & \textbf{S$_{h}$} & \textbf{2S$_{3}$} & \textbf{S$_{v}$} & \textbf{Linear} & \textbf{Quadratic} \\
         \hline
         A$_1$' & 1 & 1 & 1 & 1 & 1 & 1 & - & $x(m_x^2 - m_y^2)- 2y m_x m_y$ \\
        \hline
        A$_2$' & 1 & 1 & -1 & 1 & 1 & -1 & - & $y(m_x^2 - m_y^2) + 2x m_x m_y$ \\
        \hline
        E' & 2 & -1 & 0 & 2 & -1 & 0 & - & $x(m_x^2 - m_y^2) + 2 y m_x m_y$, $y(m_x^2 - m_y^2) -2 x m_x m_y$ \\
        \hline
        A$_1$'' & 1 & 1 & 1 & -1 & -1 & -1 & $x m_x + y m_y$, $z m_z$ & - \\
        \hline
        A$_2$'' & 1 & 1 & -1 & -1 & -1 & 1 & $y m_x - x m_y$ & - \\
        \hline
        E'' & 2 & -1 & 0 & -2 & 1 & 0 & $x m_x - y m_y$, $y m_x + x m_y$ & -
    \end{tabular}
    \caption{Irreducible representations.}
    \label{tab2}
\end{table}

Finally, by combining the linear function with cubic terms and the quadratic terms with quadratic ones, we get

\begin{align}
    E_M &= D_0 [\nabla_x (m_x^2 - m_y^2 )-2\nabla_y (m_x m_y )]+D_{\parallel} m_x (m_x^2-3m_y^2 ) \Vec{\nabla} \cdot \Vec{m} \nonumber \\
    &\quad + D_{\perp} m_y (3m_x^2-m_y^2 )(\Vec{z} \times \Vec{\nabla}) \cdot \Vec{m} + D_{\perp}^3 \nabla_x (\nabla_x^2-3\nabla_y^2 ) m_z^2
\end{align}

The first term was obtained by Laref et al. \cite{Laref2020}, the second and third terms were obtained by Ado et al. \cite{Ado2021}, the fourth term is new. Notice that the first and third terms are expected to be inactive in the bulk (in fact, we will see that the third term has the same impact as the second term in the bulk).
We use the same procedure to compute the current-driven anisotropy, by considering the appropriate combinations of the electric field $\Vec{E} = (E_x,E_y )$ with the magnetization. Similarly, we obtain

\begin{align}
    E_\alpha &= J_0 [ E_x (m_x^2 - m_y^2) - 2 E_y(m_x m_y)] + J_{\parallel} m_x (m_x^2 - 3m_y^2) \Vec{E} \cdot \Vec{m} \nonumber \\
    &\quad J_{\perp} m_y (3 m_x^2 - m_y^2) (\Vec{z} \times \Vec{E}) \cdot \Vec{m}
\end{align}

So, finally

\begin{align}
    E_M &= A \big[(\nabla_x \Vec{m})^2 + (\nabla_y \Vec{m})^2 \big] + D_0 \big[\nabla_x (m_x^2 - m_y^2) - 2 \nabla_y(m_x m_y) \big] + D_{\parallel} m_x ( m_x^2 - 3m_y^2) \Vec{\nabla} \cdot \Vec{m} \nonumber \\
    &\quad + D_{\perp} m_y (3m_x^2 - m_y^2)(\Vec{z} \times \Vec{\nabla}) \cdot \Vec{m} + D_{\perp}^3 \nabla_x (\nabla_x^2 - 3\nabla_y^2)m_z^2 \nonumber \\
    &\quad + J_0 \big[E_x(m_x^2 - m_y^2) - 2E_y(m_x m_y) \big] + J_{\parallel} m_x (m_x^2 - 3m_y^2)\Vec{E} \cdot \Vec{m} \nonumber \\
    &\quad + J_{\perp} m_y (3m_x^2 - m_y^2)(\Vec{z} \times \Vec{E}) \cdot \Vec{m} + K_0 m_z^2 + K_2 m_z^4 + K_4 m_z^6 \nonumber \\
    &\quad + K_6 \big[m_x^2(m_x^2 - 3m_y^2)^2 - m_y^2(3m_x^2 - m_y^2)^2 \big]
\end{align}

To assess the influence of these different terms on the magnetic texture, we consider a spin spiral \cite{Ado2021}

\begin{equation}
    \Vec{m} = \Vec{e}_m \cos{\alpha} + \Vec{e}_\theta \cos{(\Vec{q} \cdot \Vec{r})}\sin{\alpha} + \Vec{e}_\phi \sin{(\Vec{q} \cdot \Vec{r})}\sin{\alpha}
\end{equation}

$\Vec{e}_m$ is the direction around which the magnetization precesses with an angle $\alpha$ and a wave vector $\Vec{q}$. $\alpha = 0$, $\pi$ corresponds to uniform magnetization. Notice that there are three spin spirals of interest: the planar Néel spiral $\big(\theta = 0, \phi = 0, \alpha = \frac{\pi}{2} \big)$, the out-of-plane Néel spiral $\big(\theta = \frac{\pi}{2}, \phi = \phi_0, \alpha = \frac{\pi}{2} \big)$ propagating along $\Vec{q} = q (-\sin{\phi_0}, \cos{\phi_0})$, and the out-of-plane Bloch spin spiral $\big(\theta = \frac{\pi}{2}, \phi = \phi_0, \alpha = \frac{\pi}{2} \big)$ propagating along $\Vec{q} = q (\cos{\phi_0}, \sin{\phi_0})$.

Neglecting higher order anisotropy, the total energy functional reads

\begin{align}
    E_M &= \int d^2 \Vec{r} \bigg( A \big[(\nabla_x \Vec{m})^2 + (\nabla_y \Vec{m})^2 \big] + D_0 [\nabla_x (m_x^2 - m_y^2) - 2\nabla_y(m_x m_y)] \nonumber \\
    &\quad + D_{\parallel} m_x (m_x^2 - 3m_y^2) \Vec{\nabla} \cdot \Vec{m} + D_{\perp} m_y (3m_x^2 - m_y^2)(\Vec{z} \times \Vec{\nabla}) \cdot \Vec{m} \nonumber \\
    &\quad + D_{\perp}^3 \nabla_x (\nabla_x^2 - 3\nabla_y^2)m_z^2 + J_0 [E_x (m_x^2 - m_y^2) - 2E_y(m_x m_y)] \nonumber \\
    &\quad + J_{\parallel} m_x (m_x^2 - 3m_y^2) \Vec{E} \cdot \Vec{m} + J_{\perp} m_y (3m_x^2 - m_y^2)(\Vec{z} \times \Vec{E}) \cdot \Vec{m} + K_0 m_z^2 \bigg)
\end{align}

After injecting the spin spiral into this energy functional, we obtain

\begin{align}
    \frac{E_M}{V} &= A[q_x^2 + q_y^2] \sin{\alpha}^2 \nonumber \\
    &\quad + \frac{3}{16}(D_{\parallel} + D_{\perp})^2 (3+5\cos{2\alpha})^2 \cos{\theta}^2 \sin{\theta}^4 \sin{\alpha}^2 \nonumber \\
    &\quad + \frac{K_0}{8} (3 + \cos{2\theta} + \cos{2\alpha}(1 + 3\cos{2\theta})) \nonumber \\
    &\quad + \frac{J_0}{8}(1 + 3\cos{2\alpha}\cos(2\phi + \theta_E)\sin{\theta}^2 + (J_{\parallel} + J_{\perp})f(\theta,\phi,\alpha)
\end{align}

We do not intend to provide the general form of $f(\theta,\phi,\alpha)$ at this stage. The ground state of the system is obtained by minimizing the total energy with respect to $\Vec{q}$, which gives

\begin{equation}
    \Vec{q} = -\frac{1}{2A}\frac{3}{16}(D_{\parallel} + D_{\perp})(3 + 5\cos{2\alpha})\cos{\theta}\sin{\theta}^2 
    \begin{pmatrix}
        \sin{2\phi}\\
        \cos{2\phi}
    \end{pmatrix}
\end{equation}

Giving the total energy

\begin{align}
    \frac{E_M}{V} &= -\frac{1}{4A} \bigg(\frac{3}{16} \bigg) (D_{\parallel} + D_{\perp})^2 (3 + 5\cos{2\alpha})^2 \cos{\theta}^2 \sin{\theta}^4 \sin{\alpha}^2 \nonumber \\
    &\quad + \frac{K_0}{2} (\sin{\theta}^2 \sin{\alpha}^2 + 2\cos{\alpha}^2 \cos{\theta}^2) \nonumber \\
    &\quad + \frac{J_0}{4} \cos{(2\phi + \theta_E)}(1 + 3\cos{2\alpha})\sin{\theta}^2 + (J_{\parallel} + J_{\perp}) \cos{(4\phi_0 - \theta_E)}
\end{align}

The ground state is then obtained by minimizing the energy with respect to $\theta, \phi, \alpha$, as done in Ado et al \cite{Ado2021}. This would require further study. Instead of doing that, let’s simply inject the three different spirals into the magnetic energy.

\textbf{In-plane Néel spiral} $\big(\theta = 0, \phi = 0, \alpha = \frac{\pi}{2} \big)$

\begin{equation}
    \frac{E_M}{V} = Aq^2
\end{equation}

\textbf{Out-of-plane Néel and Bloch spirals} $\big(\theta = \frac{\pi}{2}, \phi = \phi_0, \alpha = \frac{\pi}{2} \big)$

\begin{equation}
    \frac{E_M}{V} = A q^2 + \frac{K_0}{2} - \bigg(\frac{J_0}{2} + \frac{3}{16}(J_{\parallel} + J_{\perp}) \bigg) \cos{(2\phi_0 + \theta_E)} + \frac{3}{16}(J_{\parallel} + J_{\perp}) \cos{(4\phi_0 -\theta_E)}
\end{equation}

Injecting the spin spirals into the expression for magnetic energy yields two expressions, neither of which include the expected chiral terms associated with SOT. Thus we conclude the SOT generated in an ideal FGT monolayer should not lead to domain wall motion; additional unknown mechanisms are required, as discussed above.

\end{document}